\UseRawInputEncoding
\documentclass[aps,pra,preprint,titlepage]{revtex4-2}
\usepackage{graphicx}
\usepackage{amsmath,amssymb}
\usepackage{color,soul} % for highlighting
\usepackage{hyperref}
\hypersetup{
colorlinks = true, 
urlcolor = blue, 
linkcolor = blue, 
citecolor = blue
}
\newcommand{\Zeff}{^1\!Z_\text{eff}}

\begin{document}
\setlength{\baselineskip}{0.8cm}

\title{Dissociative positronium attachment in halogen gases}

\author{R. S. Wilde$^1$} 

\author{G. F. Gribakin$^2$}

\author{I. I. Fabrikant$^3$}

\affiliation{$^1$Department of Natural Sciences, Oregon Institute of Technology, Klamath Falls, OR 97601, USA}

\affiliation{$^2$School of Mathematics and Physics, Queens University Belfast, Belfast BT7 1NN, UK}

\affiliation{$^3$Department of Physics and Astronomy, University of Nebraska,
Lincoln, Nebraska 68588-0299, USA}

\date{\today}

\begin{abstract}
We suggest that the observed large annihilation rates of ortho-positronium ($o$-Ps) in halogen gases
are due to the process of dissociative Ps attachment, ${\rm Ps} + X_2 \to {\rm Ps}X + X$,
where $X$ stands for a halogen atom. This process is similar to dissociative electron attachment which leads to formation of negative ions.
We calculate the cross section and rate of this process for the F$_2$ molecule,
for which it is exothermic, and therefore, can occur at room temperature. We start with the Ps-F$_2$ scattering calculations which 
take into account electron exchange and correlations within the framework of the free-electron-gas model. The calculations 
reveal several resonances. Similar to the process of dissociative electron attachment, a $\Sigma_u$ resonance contributes to the 
dissociative Ps attachment at thermal energies. We determine the resonance position and width as functions of the internuclear separation, 
and use them as inputs for the local version of the quasiclassical theory of dissociative attachment.  
Our calculations yield an anomalously large rate constant for the $o$-Ps annihilation process which is only one order of magnitude lower than those observed for Br$_2$ and I$_2$. 

\end{abstract}

\maketitle

\section{Introduction}

In this paper we consider the process of dissociative attachment of positronium (Ps) to halogen molecules, which may explain anomalously large annihilation rates (``quenching'') of Ps in these gases.

When fast positrons (e.g., those produced in $\beta ^+$ decay) thermalise and ultimately annihilate in matter, a sizeable fraction of them forms Ps. This is true for positrons in astrophysical environments \cite{Churazov2005,Jean2006}, in gases at normal conditions \cite{Charlton1985}, and for positrons in many condensed matter systems \cite{Schultz1988}. Ps is formed by the positron picking up an atomic or molecular electron. Its formation is typically statistical, with 25\% of it being para-Ps ($p$-Ps, total spin $S=0$), and 75\% being ortho-Ps ($o$-Ps, $S=1$). In vacuum, they annihilate predominantly by $2\gamma $ ($p$-Ps) or $3\gamma$ ($o$-Ps) annihilation, with the lifetimes of 0.125 and 142~ns, respectively. Ps is a fundamental leptonic system for testing QED \cite{Karshenboim2005} probing the effect of gravity on antimatter \cite{Cassidy2018}, and a prospective resource for antihydrogen production (see \cite{Charlton2017} and other papers in this special issue). It also has a number of important applications.

In gases and condensed-matter systems, the lifetime of $o$-Ps is reduced by its interaction with surrounding molecules or surfaces. In positron annihilation lifetime spectroscopy (PALS), this effect enables one to estimate the sizes of pores in insulators or free space in polymers. When $o$-Ps formed inside a pore collides with its walls, the positron can annihilate rapidly by $2\gamma $-emission on one of the surface electrons. The reduction in the lifetime of $o$-Ps can be related to the frequency of such collisions and the dimension of the pores using the Tao-Eldrup model (see \cite{Wada2013} and references therein). A similar processes in gases is known as the pickoff annihilation (or pickoff quenching of $o$-Ps). The total annihilation rate of $o$-Ps can then be written as
\begin{equation}\label{eq:lambda_oPs}
\lambda _{o\text{-Ps}}=\lambda _{3\gamma }+\lambda _c(n),
\end{equation}
where $\lambda _{3\gamma }$ is the intrinsic annihilation rate of $o$-Ps in vacuum, and $\lambda _c(n)$ is the annihilation rate due to Ps collisions with the gas molecules, which  depends on the number density of the gas $n$. This rate is proportional to $n$ and is traditionally parameterized by \cite{Fraser1966}
\begin{equation}\label{eq:lambda_c}
\lambda _c(n)=4\pi r_0^2cn\,\Zeff ,
\end{equation}
where $4\pi r_0^2cn$ is the Dirac $2\gamma $-annihilation rate for a positron in an uncorrelated gas of electrons (assumed to be in the $S=0$ state with the positron),
$r_0$ and $c$ are the classical electron radius and speed of light, respectively. The dimensionless parameter $\Zeff$ (also denoted $_1Z_\text{eff}$) is interpreted as the effective number of electrons per gas atom or molecule, in a singlet state relative to the positron.

The values of $\Zeff$ measured for thermalized room-temperature Ps in various gases are typically quite small, e.g., for lighter noble-gas atoms, He, Ne, and Ar, one has $\Zeff= 0.125$, 0.235, and 0.314, respectively \cite{Charlton1985}. For these atoms, many-body theory calculations \cite{Green2018,Swann2023} show good agreement with the measurements. For heavier noble gas atoms (Kr and Xe), original measurements yielded significantly larger values, $\Zeff= 0.478$ and 1.26, respectively \cite{Charlton1985}. At the same time, both gases yielded unexpectedly low Ps-formation fractions. Mitroy and Novikov \cite{Mitroy2003} showed that these findings could be explained by $o$-Ps-to-$p$-Ps conversion due to the effect of the spin-orbit interaction in Ps-atom scattering. The majority of $o$-Ps would then annihilate prior to thermalization, leading to reduced Ps-formation fractions and overestimated pickoff annihilation rates. Subsequent measurements \cite{Saito2006} confirmed this understanding. These results showed that the $o$-Ps-$p$-Ps spin-orbit conversion rates scale approximately as $Z^4$ ($Z$ being the nuclear charge), and clarified the distinction between this effect and pickoff annihilation (see also \cite{Shibuya2013}). The resulting pickoff annihilation rates are in good accord with the many-body-theory calculations \cite{Swann2023}.

The experimental room-temperature $\Zeff$ values for most molecular gases studied so far are similar to those of the noble gases. They range from $\sim$0.3 for N$_2$ and CO, to $\sim $0.5 for N$_2$O, CO$_2$, CH$_4$ and CH$_3$F, and $\sim $0.8 for C$_4$H$_{10}$ (butane), with other polyatomic gases, such as NH$_3$, CH$_3$Cl, CH$_3$Br, CCl$_2$F$_2$, SF$_6$, and C$_2$H$_6$, having similar $\Zeff$ \cite{Hyo09,Wada2012}. By contrast, several molecular gases display much larger $\Zeff$ values: $44\pm 3$ for O$_2$ \cite{Charlton1985}, 190 for NO, $1.15\times 10^4$ for Br$_2$, $1.26\times 10^4$ for I$_2$, and $5.7\times 10^5$ for NO$_2$ \cite{Hyo09,Chuang1974}. These numbers indicate that $o$-Ps annihilation in these gases is due to processes other than the simple pickoff annihilation. Thus, both O$_2$ and NO contain electrons with unpaired spins, which allows $o$-Ps to convert to $p$-Ps by electron exchange (hence, \textit{exchange quenching} \cite{Charlton1985} or \textit{spin conversion} \cite{Shibuya2013}). When separated from the spin conversion, the true pickoff annihilation for O$_2$ is estimated to give $\Zeff =0.6\pm 0.4$ \cite{Shinohara2001}, similar to those for other molecules.

Much higher $\Zeff$ values for Br$_2$, I$_2$, and NO$_2$, were interpreted as ``chemical quenching'' \cite{Chuang1974,Charlton1985}, implying formation of a Ps-molecule complex. It is known that Ps can bind to many open-shell atoms, e.g., H, Li, C, O, Na, K, and Cu, and to the halogen atoms: F, Cl, Br, and I \cite{Mitroy2002}. Ps binding is facilitated by the ability of these atoms to form stable anions, which means that the Ps-atom bound state has a strong component of the Coulomb-bound positron-anion complex. However, with the exception of Ref.~\cite{Bressanini1998} which calculated Ps binding to the OH, CH and NH$_2$ radicals, nothing is known about the possibility of Ps to attach to neutral molecules \cite{note}. It is this problem that we address in the present work. In particular, we show that low-energy dissociative Ps attachment to the halogen molecules is likely responsible for the anomalously high $\Zeff$ values observed for Br$_2$ and I$_2$.

Table \ref{tb:DE_PsA} shows the dissociation energies of the halogen molecules \cite{DissEnData,CRC} and Ps affinities (PsA) of the corresponding atoms, i.e., the Ps 
binding energies of the Ps-atom complexes. The Ps affinities are from the multi-reference configuration-interaction calculations \cite{Saito2005} 
and many-body theory calculations \cite{Ludlow2010}. In spite of some uncertainties, these values show that the process of dissociative Ps attachment (DPsA)
\begin{equation}\label{DPsA}
\text{Ps}+AB\to {\rm Ps}A + B,
\end{equation}
is exothermic for fluorine and, possibly, for bromine and iodine, and mildly endothermic for chlorine. This means that low-energy Ps atoms (e.g., room-temperature thermal, or those with 0.1--0.2~eV) can drive the process of molecular dissociation accompanied by formation of Ps-atom bound states. The Ps-atom complexes have lifetimes of about 0.5~ns~\cite{Ludlow2010}. This means that their formation by the long-lived $o$-Ps atoms will lead to rapid annihilation and result in significant increases of the $\Zeff$ parameter. We thus argue that the high $\Zeff $ values measured for Br$_2$, I$_2$ (and, possibly, for NO$_2$ too) can be explained by DPsA. Such explanation, as opposed to some rapid $o$-Ps--$p$-Ps conversion, is supported by measurements of the momentum distribution of annihilation radiation from the quenching \cite{Chuang1974}, which indicate that positrons annihilate with high-momentum atomic electrons.

\begin{table}[ht!]
\begin{ruledtabular}
\caption{Dissociation energies ($D_0$) of halogen molecules and Ps affinities (PsA) of halogen atoms.}
\label{tb:DE_PsA}
\begin{tabular}{cccc}
Species  &  $D_0$ (eV) & \multicolumn{2}{c}{PsA (eV)} \\
\cline{3-4}
  & Ref.~\cite{DissEnData,CRC} &   Ref.~\cite{Saito2005} & Ref.~\cite{Ludlow2010} \\
\hline
F$_2$      & 1.60 & 2.806 & 2.718 \\
Cl$_2$     & 2.48 & 2.350 & 2.245 \\
Br$_2$     & 1.97 & 2.061 & 1.873 \\
I$_2$      & 1.54 & 1.714 & 1.393
\end{tabular}  
\end{ruledtabular}
\end{table}

The DPsA process, Eq.~(\ref{DPsA}), is analogous to dissociative electron attachment (DEA),
\begin{equation}\label{DEA}
 e^-+AB\to A^- + B, 
\end{equation}
which leads to formation of negative ions and is important in many contexts, e.g., gas discharges, plasmas, in biological systems, and in astrophysical environments \cite{Fabrikant2017}. This and other electron-molecule processes are often mediated by creation of temporary molecular anions (here, $AB^-$), that enable efficient coupling between the light (electron) and heavy (nuclear motion) degrees of freedom.

A similar coupling is essential for positron annihilation in most polyatomic molecular gases, \begin{equation}\label{pos_ann}
 e^+ + M\to e^+M\to M^+ +2\gamma .
\end{equation}
Here annihilation proceeds via positron capture in vibrational Feshbach resonances, i.e., vibrationally excited states of the positron-molecule complex $e^+M$, which increase the positron annihilation rates dramatically \cite{Gri10}. For positron annihilation in gases, the annihilation rate is traditionally parameterised as $\lambda =\pi r_0^2cnZ_\text{eff}$, where $Z_\text{eff}$ is the effective number of target electrons that contribute to annihilation [cf. Eq.~(\ref{eq:lambda_c})]. For small molecules, such as N$_2$, O$_2$, or CO$_2$, its values for room-temperature positrons are comparable to the actual number of electrons ($Z_\text{eff}=30.8$, 26.5, and 53, respectively \cite{Charlton1985,Charlton2013}, showing some enhancement due to long-range positron-molecule attraction). However, for polyatomic molecules capable of binding the positron, values of $Z_\text{eff}$ are increased by orders of magnitude, e.g., $Z_\text{eff}=3.5\times 10^3$ for propane (C$_3$H$_8$), rising to $1.78\times 10^6$ for dodecane (C$_{12}$H$_{26}$) \cite{Iwata95}. For larger polyatomics the positron can remain attached to the molecule for longer due to intramolecular vibrational energy redistribution \cite{Uze91}. As a result, the lifetime of the complex against positron autodetachment becomes large and the annihilation cross section increases and can become comparable to the geometrical cross section of the molecule.

Similar phenomena can occur in resonant Ps scattering. Moreover, when the DPsA channel is open, this process would also strongly increase the annihilation rate. To affect the annihilation rate at thermal energies, the reaction (\ref{DPsA}) should be exothermic, with the reaction threshold
\[
E_\text{th}=D_0-{\rm PsA},
\]
being negative (or mildly endothermic, with a small positive $E_\text{th}$). According to the data in Table \ref{tb:DE_PsA}, the reaction involving F$_2$ is most certainly exothermic. Calculations for the lightest halogen molecule are also less challenging. Hence, in this paper we present theoretical results for Ps interaction with F$_2$ and show that DPsA indeed leads to an anomalously high annihilation rate with $\Zeff \sim 10^3$. Atomic units (a.u.) are used throughout the paper unless stated otherwise.

\section{Electron and ${\bf Ps}$ scattering from F$_2$ and potential energy curves for F$_2^-$ and ${\bf PsF}_2$}

Our DPsA calculations follow the general method used for DEA mediated by a temporary anion state (see, e.g., Ref.~\cite{Fabrikant2016}). To implement this, we first need to obtain the resonance position and width for a range of internuclear separations. This can be done by calculation of Ps-F$_2$ collisions.
The problem of low-energy Ps scattering by molecules is very challenging theoretically because of the importance of exchange and correlations in these collisions. Here we use the approach of Ref.~\cite{PsFEG} that accounts for these effects using the free-electron-gas (FEG) approximation. It enables one to calculate the Ps-target potential and was shown to work well for Ps scattering from the noble gases, as well as N$_2$, CO$_2$, and O$_2$ (see below).

To check the reliability of the model, we applied it first to electron-F$_2$ scattering in $^2\Sigma_u$ symmetry. The resonance of this symmetry is responsible for DEA to F$_2$ at low energies. Several calculations of this resonance and corresponding DEA cross sections are available in the
literature \cite{Rescigno1976,Hall1978,Hazi1981,Bardsley1983,Kalin1990,Ingr1999,Brems2002,Honigmann2012,Shuman2013,Fabrikant2016}. Analysis of experimental data on the attachment rate coefficients \cite{Shuman2013} showed that the best theoretical DEA
cross sections were generated by Hazi {\it et al.} \cite{Hazi1981} using the Stiltjes momentum imaging technique for calculation of the resonance 
width. In a more recent study \cite{Fabrikant2016}, we used the resonance position and width from $R$-matrix calculations. The corresponding width turned out to be a factor of two greater than that of Hazi {\it et al.}, while the anion potential energy curve was consistent with other accurate calculations.

Our present fixed-nuclei treatment of $e^-$-F$_2$ and Ps-F$_2$ systems uses the scattering potentials calculated in the FEG approximation \cite{Hara67,PsFEG}.  In both cases, the FEG potentials depend on the ground-state electron density which was obtained from the PySCF suite of quantum chemistry codes using the cc-pVTZ basis \cite{pyscf1,pyscf2,pyscf3}. 

We have previously applied the FEG model to Ps scattering by rare-gas atoms \cite{Ps_rg2}, molecular targets such as N$_2$, O$_2$,CO$_2$ 
\cite{PsN2,res_PsN2,PsO2_CO2,Ps_O2} as well as polar molecules \cite{Ps_polar}. These calculations corroborated the similarity between the electron and Ps total scattering cross sections at equal projectile velocities for nonpolar molecules, as seen in experiment above the Ps ionization threshold \cite{Brawley_2010}.  The calculations also confirmed observations of Ps scattering resonances near the Ps ionization threshold \cite{Psres_2010,Psres_2017}.

The long-range behavior of the scattering potentials is different for electron scattering compared with Ps scattering.  In the former case (e$^-$-F$_2$), the polarization of the molecule dominates the interaction,
$V(r)\simeq -\alpha /2r^4$. We describe it using the spherical polarizability $\alpha = 7.84$~a.u. \cite{alpha_F2} of F$_2$ at the equilibrium internuclear separation of $R_e=2.668$~a.u.  In the Ps-F$_2$ case, the long-range interaction is of the van der Waals type, $V(r)\simeq -C_6/r^6$, and we use $C_6=73.84$~a.u., obtained from the London formula \cite{London} at equilibrium. This difference in the long range behavior of the scattering potentials leads to different behavior between electron and Ps scattering at low energies (velocities) \cite{Ps_rg1}.

\begin{figure}
%\picturelandscape{0}{elasticXC.pdf}
\includegraphics[scale=.7]{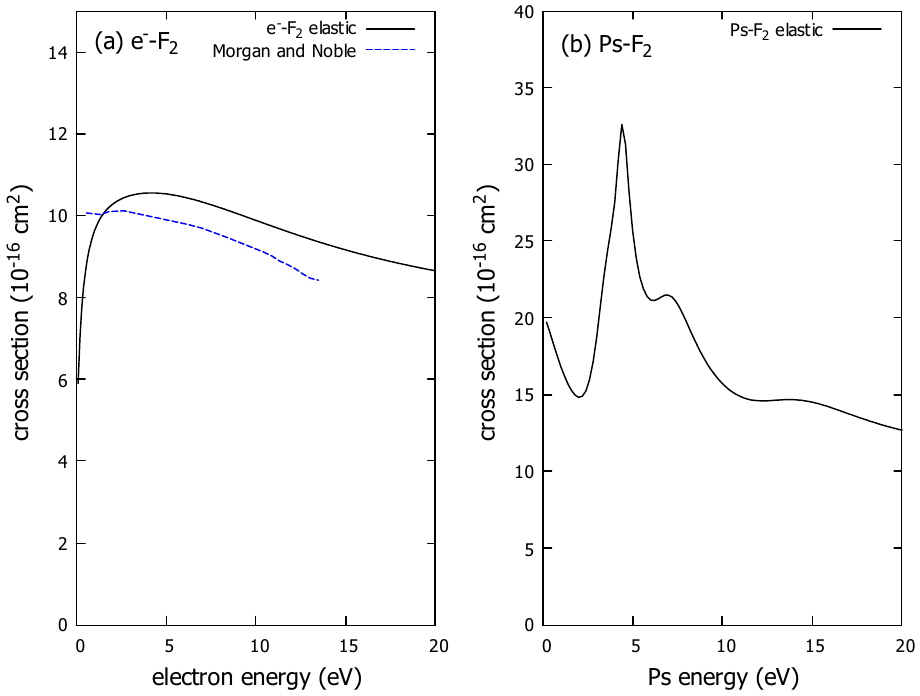}
\caption{(Color online) (a) Total elastic cross section for e$^-$-F$_2$. Solid line: present FEG results, blue dashed line: R-matrix results of Morgan and Noble \cite{Morgan_Noble}. (b) Present FEG total elastic cross section for Ps-F$_2$.}
\label{fig:elasticXC}
\end{figure}

Figure~\ref{fig:elasticXC} shows the elastic cross sections for the electron and Ps scattering from F$_2$ at the equilibrium nuclear separation.  The present e$^-$-F$_2$ cross section is in good agreement with the $R$-matrix calculations of Morgan and Noble \cite{Morgan_Noble}. The electron cross section does not display any resonance structure, whereas in the Ps-F$_2$ case, we see several resonances similar to those seen for Ps scattering from N$_2$, O$_2$, and CO$_2$.

\begin{figure}
%\picturelandscape{0}{partialXC.pdf}
\includegraphics[scale=.7]{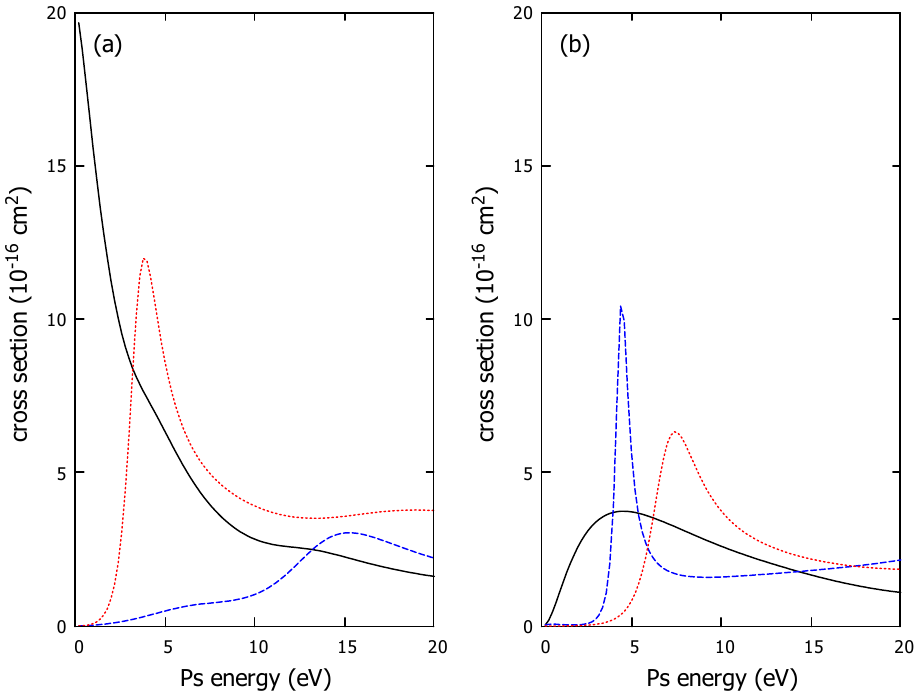}
\caption{(Color online) (a) Partial elastic cross section for Ps-F$_2$ scattering in gerade symmetries. Solid line: $\Sigma_g$, blue dashed line: 
$\Pi_g$, red dotted line: $\Delta_g$. (b) Partial elastic cross section for Ps-F$_2$ scattering in ungerade symmetries. Solid line: $\Sigma_u$, blue dashed line: $\Pi_u$, red dotted line: $\Delta_u$.
}
\label{fig:partialXC}
\end{figure}

\begin{figure}
%\picturelandscape{0}{eigsumDgPu.pdf}
\includegraphics[scale=.7]{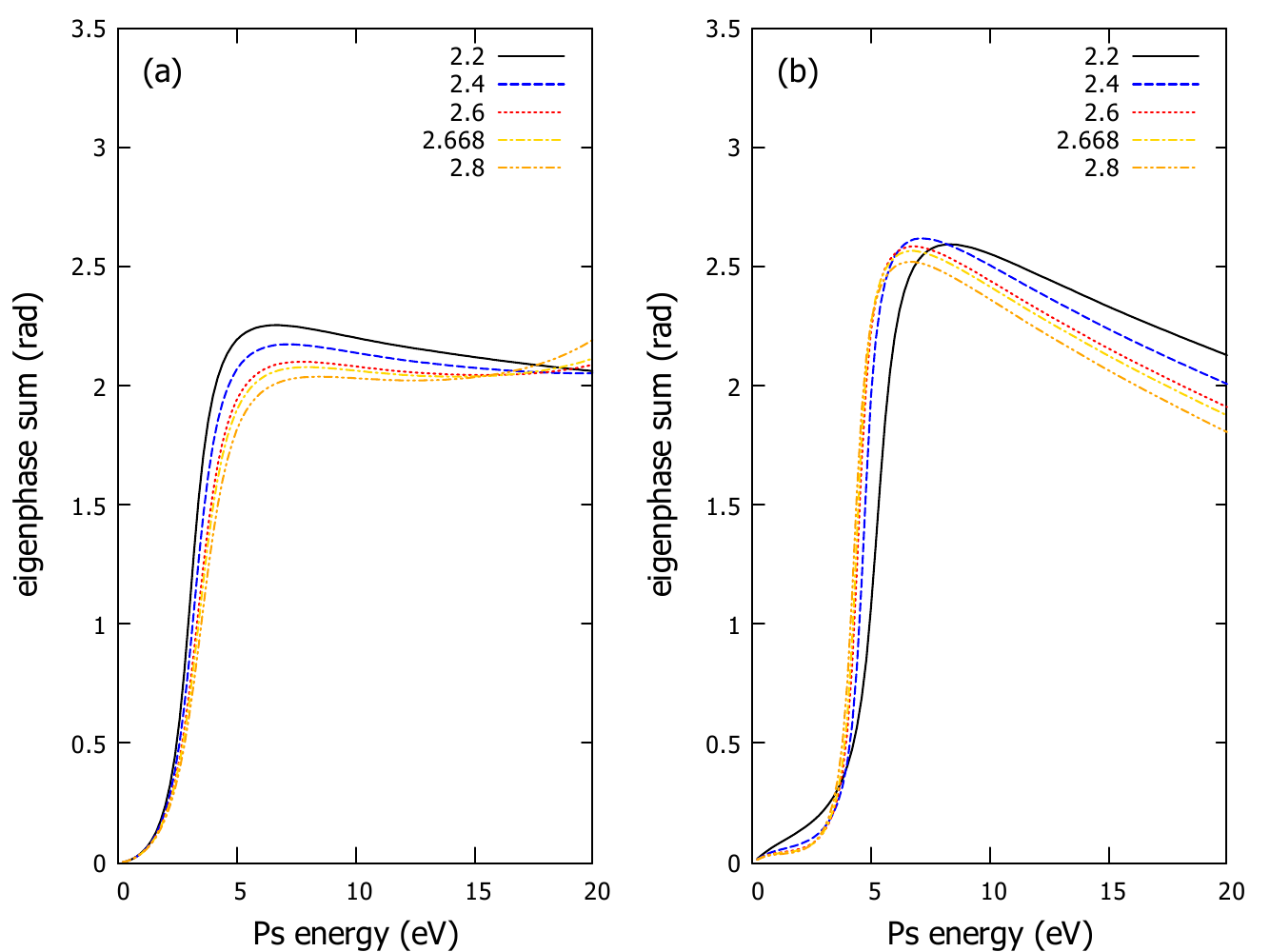}
\caption{(Color online) Ps-F$_2$ eigenphase sum for the $\Delta_g$ symmetry (a) and $\Pi_u$ symmetry (b) at several values of the internuclear separation $R$ in a.u.}
\label{fig:eigsumDgPu}
\end{figure}

Figure \ref{fig:partialXC} shows the partial $\Sigma $, $\Pi$ and $\Delta $ cross sections with gerade and ungerade symmetries for Ps-F$_2$ elastic scattering at the equilibrium separation. The cross sections display resonances in several symmetries, with the lowest two being $\Delta_g$ at 3.4~eV and $\Pi_u$ at 4.4~eV. 
These resonances, as well as the two at higher Ps energies ($\Delta_u$ and $\Pi_g$), do not
contribute to DPsA because the corresponding potential energy curves do not cross the neutral curve.  
In figure \ref{fig:eigsumDgPu} we show the eigenphase sums for the $\Delta_g$ and $\Pi_u$ symmetries for several values of the internuclear separation. Their position remains well above zero energy even at $R=2.8$~a.u.
%added on 5/15
The absence of similar resonances in $e-$F$_2$ scattering is in variance with previously experimentally observed \cite{Brawley_2010,Psres_2010}
and theoretically confirmed  \cite{Ps_rg1,PsN2,PsO2_CO2}
similarities 
between electron and Ps scattering when viewed as functions of the projectile velocity $V$ for $V\sim 1$ a.u. and higher, so for Ps energies higher 
than 6-7 eV. The strongest resonance features we see in Ps-F$_2$ scattering cross section are mostly below 
this energy, therefore the absence of these in electron scattering should not be surprising. This situation is similar to that observed in scattering 
by the O$_2$ molecule \cite{Ps_O2} whereby the low-energy resonance was theoretically predicted for Ps scattering, but not for electron scattering. 
In the case of F$_2$, from comparison of exchange and correlation potentials for two projectiles, we conclude that the total potentials with the centrifugal barrier 
added are quite similar, the one for Ps being a bit shallower but broader. The effect of the mass of Ps likely comes into play here, making it 
easier to create resonances.

\begin{figure}
%\picturelandscape{0}{eigsumSu.pdf}
\includegraphics[scale=.7]{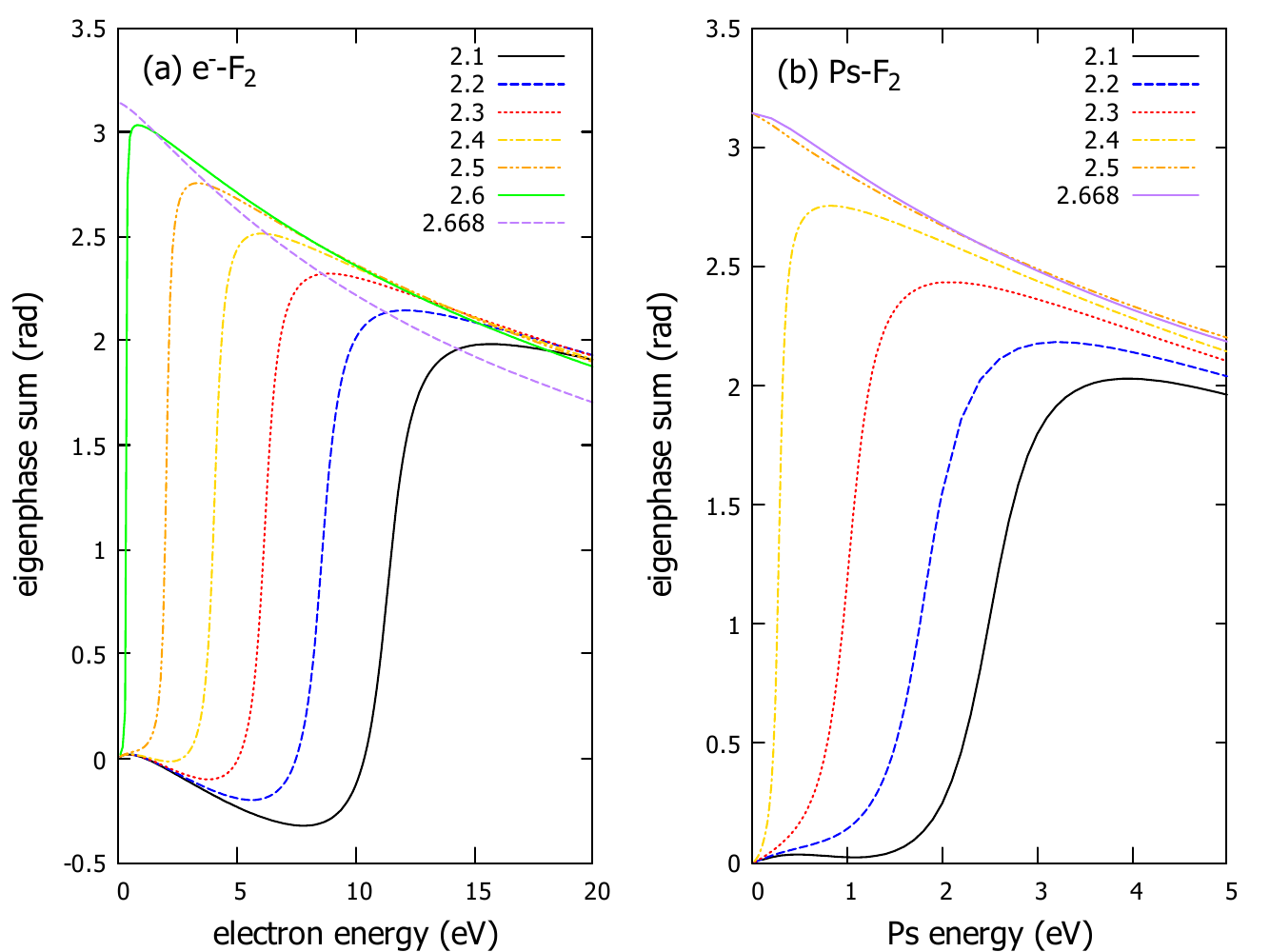}
\caption{(Color online) e$^-$-F$_2$ eigenphase sum for the $\Sigma_u$ symmetry (a) and Ps-F$_2$ eigenphase sum for the $\Sigma_u$ symmetry (b) at several values of the internuclear separation $R$ in a.u. 
} \label{fig:eigsumSu}
\end{figure}

At low (thermal) energies, only the $\Sigma_u$ resonance contributes to DEA to F$_2$.  In a similar way, the $\Sigma_u$ resonance is also responsible for DPsA at thermal energies. Figure~\ref{fig:eigsumSu} shows the $\Sigma_u$ phase shifts for the electron and Ps 
scattering by F$_2$ for a range of internuclear separations $R$. In both cases, the resonance becomes a true bound state for $R\approx R_e$.

%\GGcomm{[GG: SHOULD THE PHASE SHIFTS FOR $R=2.668$~a.u. for Ps-F$_2$ and $R=2.5$ and %2.668~a.u. for $e^-$-F$_2$ BE SHIFTED UP BY $\pi $? IT IS CLEAR THAT FOR THESE $R$ THE %RESONANCES BECOME BOUND STATES, SO THE PHASE SHIFTS GO TO $\pi$ INSTEAD OF 0 (Levinson's %theorem).] RW-Fig.4 updated with these phase shifts going to $\pi$}

\begin{figure}
\includegraphics[scale=.7]{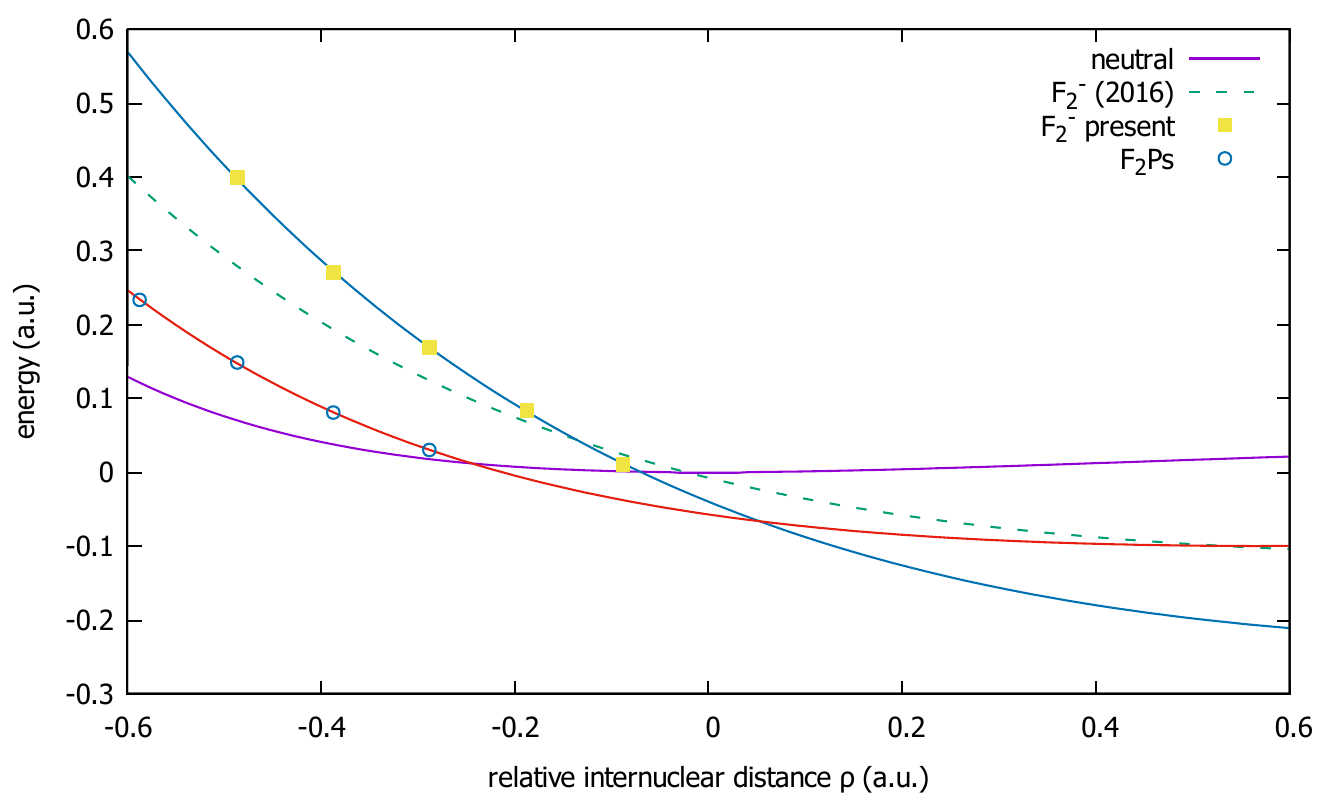}
\caption{ Potential energy curves for F$_2$ (solid purple), F$_2^-$ (solid blue and yellow squares), and PsF$_2$ (solid red and open circles) as 
functions of $\rho=R-R_e$. The dashed curve labeled ``F$_2^-$ (2016)'' is the anion curve used in Ref. \cite{Fabrikant2016}. The solid blue and red curves joining the symbols are the results of interpolation/extrapolation using the Morse parameterization.
}
\label{fig:curves}
\end{figure}

Figure \ref{fig:curves} shows the $\Sigma _u$ potential energy curves for the $e^-$-F$_2$ (i.e., F$_2^-$) and Ps-F$_2$ systems obtained from the 
scattering calculations described above. The anion curve $U(R)$ is also compared with the previous results \cite{Fabrikant2016}. The zero-energy level corresponds to energy of the neutral molecule at $R=R_e$, and the internuclear separation is given in terms of $\rho=R-R_e$.

The present anion curve crosses the neutral curve at $\rho_{\rm cr}=R_{\rm cr}-R_e=-0.071$~a.u. compared to $-0.023$ a.u. in the earlier 
calculation \cite{Fabrikant2016}. It still lies in the Franck-Condon region for the transition from the ground vibrational state since the 
left turning point for this state
is at $\rho=-0.107$ a.u. We did not carry out energy calculations in the region $R>R_{\rm cr}$, as this is not critical for DEA. However, 
extrapolation of our results to this region with the account of $U(\infty)=-0.066$ a.u. produces a very deep minimum at $\rho=1.0$ a.u. with 
$U_\text{min}=-0.23$~a.u. This is certainly an overestimate since in more accurate calculations the value at the minimum is about $-0.16$ a.u. 
Note that the asymptotic value of $U(R)$ was obtained from the F$_2$ dissociation energy 
of $D_0=1.60$ eV \cite{DissEnData}, vibrational quantum $\hbar\omega=0.11$ eV, and the electron affinity of F, $\text{EA}=3.40$~eV.

The PsF$_2$ curve of the $\Sigma_u$ symmetry demonstrates a lower position of the resonance and a smaller value of the crossing point, $\rho_{\rm cr}=-0.250$~a.u. Therefore, it lies in the classically forbidden range outside the Franck-Condon region, and we should expect the DPsA cross section to be lower than the DEA cross section. The lower position of the resonance for Ps-F$_2$ compared to the $e^-$-F$_2$ scattering could be due to the weaker centrifugal potential which is inversely proportional to the mass of the projectile.

Table \ref{tb:width} shows the adiabatic resonance width $\Gamma (R)$ for F$_2^-$ and PsF$_2$ systems and compares it with that used in \cite{Fabrikant2016}. While the latter turned out to be an overestimate, the present calculations most certainly underestimate
the width. This should lead to a further reduction of the DEA cross section, since at low energies, the survival factor is close to unity, and the cross section is proportional to the adiabatic width [see Eq.~(\ref{eq:sigma}) below].

\begin{table}[ht!]
\begin{ruledtabular}
\caption{Adiabatic resonance width in eV for F$_2^-$ and PsF$_2$ for a range of nuclear separations $R$.}
\label{tb:width}
\begin{tabular}{cccc}
 & \multicolumn{3}{c}{$\Gamma (R)$ (eV)} \\
\cline{2-4}
$R$ (a.u.) & F$_2^-$, present &   F$_2^-$, Ref.~\cite{Fabrikant2016} & PsF$_2$, present \\
\hline
 2.1        &         1.6270 &   $-$           &          0.9577 \\
 2.2        &         1.2826 &    $-$         &          0.7743 \\
 2.3        &         0.9180 &    $-$          &          0.4513 \\
 2.4        &         0.6063 &   1.786        &          0.0954 \\
 2.5        &         0.2897 &   0.912        &             0   \\
 2.6        &         0.0503 &   0.247        &             0   
\end{tabular}  
\end{ruledtabular}
\end{table}

\section{Dissociative electron attachment to F$_2$}

In \cite{Fabrikant2016} we compared the results of local and nonlocal calculations of DEA to F$_2$. The nonlocal effects are important in this case 
since the local theory exhibits several unphysical features. In addition to the incorrect threshold behavior (due to the width $\Gamma (R)$ being 
independent of the electron energy), the local theory for F$_2$ also produces unphysical oscillations in the cross section as a function of energy 
due to the oscillatory behavior of the survival factor. However, in the quasiclassical (WKB) version of the DEA theory \cite{Kalin1990,Elets1984}, these features disappear, and, in particular,
the cross section exhibits the correct threshold behavior. To understand this we note that  in the WKB approximation, the electron capture occurs at the fixed internuclear separation, or the Franck-Condon point $R=R_F$, which is found from
\begin{equation}
U(R_F)-U_0(R_F)=E,
\label{FC}
\end{equation}
where $U(R)$ and $U_0(R)$ are the potential energy curves for the anion and for the neutral molecule, respectively, and $E$ is the incident electron energy \cite{Elets1984} [see Eq. (\ref{FCap}) in Appendix \ref{app}]. Near the crossing point, the adiabatic width of the local theory behaves according to the Wigner threshold law, as
\[
\Gamma(R) = {\rm const}\times [U(R)-U_0(R)]^a,
\]
where $a$ is the threshold exponent, $a=3/2$ for the $\Sigma_u$ resonance. Substituting Eq.~(\ref{FC}), we obtain the correct threshold behavior,
\[
\Gamma(E)\equiv \Gamma (R_F) = {\rm const}\times E^a.
\]
Further details of derivation of the quasiclassical expression for the DEA cross sections are given in Appendix \ref{app}. The result is given by Eq. (\ref{crs_ap}), and has the form
\begin{equation}\label{eq:sigma}
\sigma = \frac{4\pi ^2}{k^2}\Gamma (R_F)F_v(E)s,
\end{equation}
where $k$ is the electron momentum ($E=k^2/2$ in atomic units), $F_v(E)$ is the generalised Franck-Condon factor calculated using the uniform Airy function approximation \cite{Elets1984}, and $s$ is the survival factor
\[
s=\exp \left(-\int _{R_1(E)}^{R_{\rm cr}}\frac{\Gamma (R)}{v(R)} dR\right),
\]
where $R_1(E)$ is the turning point of the anion motion, $R_{\rm cr}$ is the potential curve crossing point at which $U(R_{\rm cr})=U_0(R_{\rm cr})$, and $v(R)$  is the classical velocity of the nuclei in the anion state.
The cross section in Eq.~(\ref{eq:sigma}) has the correct threshold behavior of the DEA cross section $\sigma \propto \Gamma(E)/E \propto E^{a-1}$ 
\cite{Elets1984}. In contrast, in the exact version of the local theory, all internuclear distances contribute to the DEA amplitude and cross 
section, including those for which $\Gamma$ is finite, and the cross section diverges as $1/E$. 

\begin{figure}
\includegraphics[scale=.7]{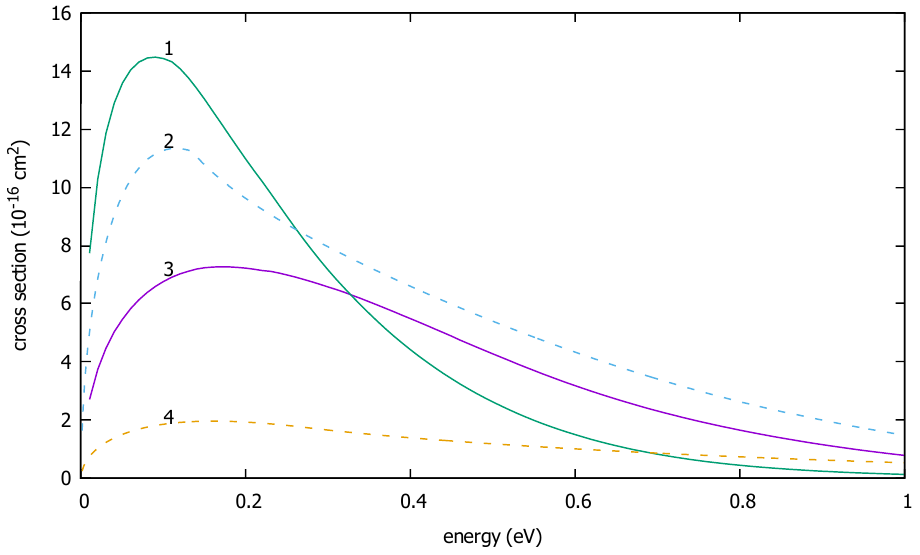}
\caption{DEA cross sections for F$_2$. Curve 1: nonlocal results calculated with the width obtained from $R$-matrix calculations \cite{Fabrikant2016}; curve 2: quasiclassical approximation of the local theory \cite{Fabrikant2016}; curve 3: nonlocal results \cite{Shuman2013} obtained with the resonance width of Hazi {\it et al.} \cite{Hazi1981};
curve 4: present quasiclassical (WKB) result, Eq.~(\ref{eq:sigma}). 
} \label{fig:F2xsec}
\end{figure}

Figure \ref{fig:F2xsec} shows the DEA cross sections for F$_2$ calculated using different methods. A comparison of the WKB version of the local theory with the nonlocal results (both from Ref.~\cite{Fabrikant2016}) shows that at low electron energies, the former underestimates the cross 
sections by about a factor of two. Calculations with the present width lead to a further reduction in magnitude. This makes our cross section comparable with the previous nonlocal calculations
\cite{Kalin1990,Brems2002} where the cross sections are too small, apparently because of the underestimated resonance width. 
We conclude that the uncertainty in the width is a major factor that affects the accuracy of the cross section. Hence, at the present level of 
accuracy of calculation of the width with the FEG model, it suffices to use the WKB version of the local theory, Eq.~(\ref{eq:sigma}). The present 
DEA calculations in the WKB approximation underestimate the best results by a factor of four. We expect the same accuracy for the DPsA cross sections shown in Sec.~\ref{sec:DPsA}.

\section{Dissociative ${\bf Ps}$ attachment: cross sections and thermal rate coefficient}\label{sec:DPsA}

\begin{figure}
\includegraphics[scale=.7]{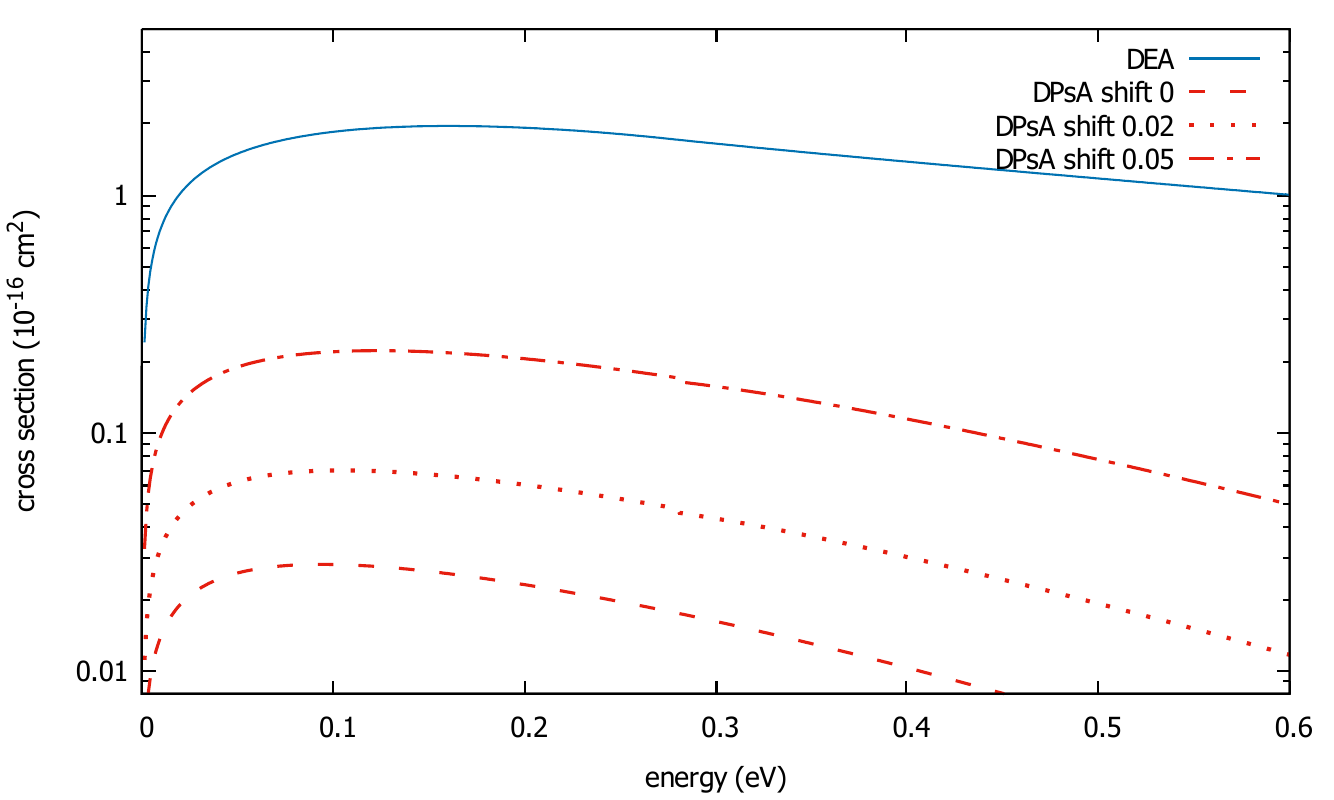}
\caption{Comparison of DEA and DPsA to the F$_2$ molecule:
Solid blue curve: present quasiclassical results for DEA; red curves: present quasiclassical results
for DPsA obtained with various horizontal shifts of the PsF$_2$ potential energy curve, see text. 
} \label{fig:DPsA}
\end{figure}

Figure \ref{fig:DPsA} presents a comparison of the DEA and DPsA cross sections for attachment to F$_2$ calculated in the WKB approximation of the local theory, Eq.~(\ref{eq:sigma}). The calculations use the  potential energy curves shown in Fig.~\ref{fig:curves} and widths $\Gamma (R)$ shown 
in Table \ref{tb:width}, and for Ps, we replace $k$ in Eq.~(\ref{eq:sigma}) by the Ps momentum $k_{\rm Ps}$, linked to its energy by $E=k_{\rm Ps}^2/4$. In addition to the calculation with the {\it ab initio} PsF$_2$ potential energy curve, we show the cross sections obtained with the original curve shifted horizontally to the right by $\Delta R=0.02$ and 0.05~a.u. (see discussion below).
The peak value of the DEA cross section (solid blue curve) is greater than that of DPsA, calculated with the original curve (dashed red curve), by a factor of 70. The reason for this is clear from a comparison of the potential energy curves in Fig. \ref{fig:curves} and resonance widths in Table \ref{tb:width}.  The curve crossing in the Ps case occurs in the classically forbidden region, which strongly reduces the Franck-Condon factor and the capture probability. The lower width and higher incident momentum (for a given energy) in the Ps case leads to a further reduction of the cross section.

In spite of the DPsA cross section being small on the atomic scale, its contribution to the o-Ps annihilation rate expressed in terms of $\Zeff$, is quite signficant. Since the process of DPsA leads to rapid annihilation, its contribution to $\Zeff$ is estimated from the ratio of the DPsA rate to the Dirac annihilation rate [cf. Eq.~(\ref{eq:lambda_c})],
 \[
 \Zeff =\frac{\langle \sigma V\rangle}{4\pi r_0^2 c}.
 \]
The angular brackets in the numerator stand for the thermal, Maxwellian average of the rate coefficient $\sigma V$, where $V=k_{\rm Ps}/2$ is the Ps velocity (in atomic units).

Calculation of the rate coefficient at room temperature for the ground vibrational state of F$_2$ gives $\langle \sigma V\rangle =0.178\times 10^{-10}$ cm$^3$/s, which corresponds to $\Zeff=596$. A further averaging over the initial vibrational-state distribution of F$_2$ at room temperature
leads to $\Zeff=629$. This reflects the fact that cross section increases dramatically with the growth of the initial vibrational level, affecting the rate coefficient even at room temperature, even though the thermal energy $k_BT$ is 4 times smaller than the vibrational quantum $\hbar \omega $. These numbers should be regarded as a lower bound since the DPsA cross section we obtained is likely an 
underestimate by as much as a factor of five. However, even this value is anomalously large: it is greater than the spin-orbit quenching 
rate coefficients for O$_2$ and NO, and is only one order of magnitude lower than the rates observed
for Br$_2$ and I$_2$ \cite{Hyo09,Wada2012}.

In order to estimate how the position of the curve crossing influences the cross section and rate coefficient, we performed additional calculations with the PsF$_2$ potential energy curve shifted to the right by small amounts (0.02 and 0.05~a.u.), assuming the same adiabatic width as a function of energy. Since the Franck-Condon factor grows exponentially when the crossing point moves towards the classically allowed region, the cross sections and rate coefficients increase dramatically.
In particular, shifting the curve by $\Delta \rho =0.02$ and $\Delta \rho =0.05$~a.u., increases Boltzmann-averaged $\Zeff$ to 1507 and 4281 respectively. The corresponding cross sections are shown in Figure \ref{fig:DPsA}, with Table \ref{tb:rates} summarizing the results for the rate coefficients and $\Zeff$. Note that in both cases, the crossing point $\rho _{\rm cr}$ remains to the left of the classically-allowed region ($\rho = -0.107$~a.u.). 

\begin{table}[ht!]
\begin{ruledtabular}
\caption{Thermal-averaged rate coefficients and $\Zeff$ for DPsA by F$_2$ at $T=300$~K with various shifts $\Delta \rho $ of the PsF$_2$ potential energy curve.}
\label{tb:rates}
\begin{tabular}{cccc}
$\Delta\rho$~(a.u.) & $\rho_{\rm cr}$~(a.u.) & Rate ($10^{-10}~{\rm cm}^3$/s) & $\Zeff$ \\
\hline
0   & $-0.2505$ & 0.1882 & 629  \\
0.02   & $-0.2213$ & 0.4509 & 1507  \\
0.05   & $-0.1810$ & 1.281 & 4281 \\
\end{tabular}  
\end{ruledtabular}
\end{table}

\section{Conclusion}
Although measurements of o-Ps annihilation rates in the F$_2$ gas are not available, the present calculations strongly suggest that anomalously large annihilation rates observed in heavy halogen gases, Br$_2$ and I$_2$, can be due to the process of DPsA. Hence, we believe that the DPsA is the likely mechanism responsible for ``chemical quenching''. Perhaps a similar mechanism is operative in the NO$_2$ gas where the annihilation rate is extremely high with $\Zeff =8\times 10^5$. This could be explained if the DPsA cross section in this gas at thermal energies is of the order of $10^{-15}$ cm$^2$.
The measured DEA cross sections for NO$_2$ \cite{Rangwala2003} give a much lower value of $\sim \!10^{-17}$ cm$^2$. The theoretical paper \cite{Liu2021} reports a much higher peak value of $\sim \!10^{-16}$ cm$^2$, but the calculated threshold for the $e^-+{\rm NO}_2\to {\rm NO}+{\rm O}^-$ reaction is 1.71~eV, making the DEA cross section negligible at thermal energies. We have seen, however, that replacing the electron by Ps can drastically change the dissociative attachment dynamics due to the
sensitivity of attachment cross sections to the potential energy curves and resonance widths. Therefore,
it is not unlikely that DPsA cross section for NO$_2$ is higher at thermal energies.

\section*{Acknowledgement}
This work has been supported by the National Science Foundation, Grant No. PHY-2309261.

\appendix

\section{Outline of quasiclassical theory of dissociative attachment}\label{app}

Since our version of the quasiclassical theory of dissociative attachment is somewhat different from the original work of Kazansky and Yelets~\cite{Elets1984}, 
we outline our derivation here.

The differential cross section, averaged over all molecular orientations, in the  axial-recoil approximation is given by \cite{Bardsley}
\[
\frac{d\sigma _v}{d\Omega_{\bf K}}=\frac{2\pi^2K}{Mk^2}|\xi_v(R,\hat{\bf K})|^2,~~\quad R\to\infty ,
\]
where {\bf K} is the momentum of the relative motion of the molecular fragments, $M$ is the reduced mass of the target, $k$ is the projectile (electron or Ps) momentum, and the function $\xi_v(R,\hat{\bf K})$ describes the fragments' relative motion. The subscript $v$ indicates the initial vibrational
state of the molecule. Expansion of $\xi_v(R,\hat{\bf K})$ in partial waves
\[
\xi_v(R,\hat{\bf K})=\sum_{lm}Y_{lm}(\hat{\bf K})\xi_{vlm}(R),
\]
gives the equation for the partial amplitude $\xi_{vlm}(R)$ in the local approximation,
\begin{equation}
\left[-\frac{\hbar^2}{2M}\frac{d^2}{dR^2}+U(R)-\frac{i}{2}\Gamma(R)-E\right]\xi_{vlm}(R)=-i^l\frac{\gamma_{lm}(R)}{\sqrt{2\pi}}\zeta_v(R) ,
\label{local_eq}
\end{equation}
where $\gamma_{lm}(R)$ is the complex partial capture amplitude, $U(R)$ the potential energy of the resonance state, $\Gamma(R)$ its width, 
and $\zeta_v(R)$ is the wave function of the initial vibrational state.
From now on we will assume for simplicity that the capture occurs into a state with a fixed angular momentum $l$ and its projection $m$, and will drop the index $m$. The solution of Eq.~(\ref{local_eq}) can be written as
\[
\xi_{vl}(R)=-\frac{i^l}{\sqrt{2\pi}}\int G(R,R')\gamma_l(R')\zeta_v(R')dR',
\]
where 
\[
G(R,R')=\frac{M}{iK}\psi^{(r)}(R_<)\psi^{(+)}(R_>),
\]
is the Green's function, $R_<=\min \{R,R'\}$, $R_>=\max \{R,R'\}$, and $\psi^{(r)}$ and $\psi^{(+)}$ are the regular and outgoing-wave solutions of the homogeneous version of Eq.~(\ref{local_eq}) with the asymptotic behavior
\[
\psi^{(+)}\simeq e^{iKR},\qquad \psi^{(r)}\simeq e^{-iKR}-{\cal S}e^{iKR},
\]
${\cal S}$ being the scattering matrix. Finding the asymptotic form of $\xi_{vl}(R)$ for $R\to \infty $, we obtain for the integrated cross section
\begin{equation}
 \sigma_v=\frac{\pi M}{Kk^2}\bigg|\int \psi^{(r)}(R)\gamma_l(R)\zeta_v(R)dR\bigg|^2. 
\label{sigma}
\end{equation}

We now use the quasiclassical approximation by representing the wave functions in the form \cite{Landau}
\[
\psi^{(r)}(R)=2\sqrt{\frac{K}{p(R)}}\sin[S(R)+\pi/4],
\]
\[
\zeta_v(R)=\sqrt{\frac{2M\omega_v}{\pi p_0(R)}}\sin[S_0(R)+\pi/4],
\]
where $p_0(R)$ and $p(R)$ are classical momenta for the nuclear motion in the neutral potential $U_0(R)$ and the anion potential
$U(R)$ respectively, $S_0(R)$ and $S(R)$ are corresponding classical actions, and $\omega_v$ is the classical frequency of the motion
in the potential $U_0(R)$. The equations above are given for the classically allowed region, but an extension to the classically 
forbidden region is straightforward. 

The integral in Eq. (\ref{sigma}) is evaluated using the saddle-point method. The result is reduced to the product of the generalized Franck-Condon 
factor and the survival probability factor. Since the derivation of the latter is well-known~\cite{Bardsley,OMalley}, we will outline here calculation 
of the former assuming that the resonance width $\Gamma(R)$ is sufficiently small so that the saddle point is given by the solution of the equation
\begin{equation}
E_t-U(R)=\epsilon_v-U_0(R),
\label{FCap}
\end{equation}
where $E_t$ is the total energy, and $\epsilon_v$ is the vibrational energy of the initial state. Equation~(\ref{FCap}) represents the Franck-Condon condition: the kinetic
energy of the nuclear motion does not change during electron capture. Denoting the solution of this equation by $R_F$
and assuming that $\gamma_l(R)$ is changing much slower than $\psi^{(r)}$ and $\zeta(R)$, we obtain
\[
\int \psi^{(r)}(R)\gamma_l(R)\zeta_v(R)dR=\gamma_l(R_F)I_v,
\]
where
\[
I_v=\int \psi^{(r)}(R)\zeta_v(R)dR =\left[\frac{2KM\omega_v}{\pi p_0(R_v)p(R_v)}\right]^{1/2}\int\cos[S(R)-S_0(R)]dR.
\]
The saddle-point method is hence reduced to the stationary phase method, with the result
\[ I_v=\left[\frac{4\omega_vK}{p_0(R_F)|F-F_0|}\right]^{1/2}\cos(\Delta S-\pi/4) \]
where $F_0=-(dU_0/dR)_{R=R_F}$, $F=-(dU/dR)_{R=R_F}$, and $\Delta S=S_0(R_v)-S(R_v)$.
Note also that according to Eq. (\ref{FC}) or (\ref{FCap}), $p_0(R_F)=p(R_F)$. 

This result is valid if the Franck-Condon point lies in the classically-allowed region sufficiently far from the turning points. Generalization to an arbitrary position of $R_v$ is achieved by the uniform Airy function approximation \cite{Elets1984}. Briefly, it is reduced to the substitution
\[
\cos(\Delta S-\pi/4)~\to ~\sqrt{\pi }\left(\frac{3}{2}\Delta S\right)^{1/6}{\rm Ai}\left[\pm\left(\frac{3}{2}\Delta S\right)^{2/3}\right],
\]
where Ai is the regular Airy function, and the sign of its argument depends on whether the Franck-Condon point lies in the 
classically forbidden or classically allowed region. We finally obtain the following expression for the cross section in the quasiclassical approximation,
\begin{equation}
 \sigma_v=\frac{4\pi^2}{k^2}s\frac{\omega_v\Gamma(R_F)}{p_0(R_F)|F-F_0|}\left(\frac{3}{2}\Delta S\right)^{1/3}{\rm Ai}^2
\left[\pm \left(\frac{3}{2}\Delta S\right)^{2/3}\right]
\label{crs_ap}
\end{equation}
where we use $\Gamma(R_F)=\sum_l|\gamma_l (R_F)|^2$ and the fact that the sum is reduced to one term if the resonance is dominated by a single partial wave $l$. 

The survival factor $s$ is given by~\cite{Bardsley,OMalley}
\[ s=\exp\left[-\int_{R_1}^{R_{\rm cr}}\frac{\Gamma(R)}{v(R)}dR\right] \]
where $v(R)=p(R)/M$ is the classical velocity for the motion in the potential $U(R)$. The lower integration limit $R_1$ is
reduced to $R_F$ if $R_F$ lies in the classically allowed region. Otherwise the turning point for the
motion in the potential $U(R)$ should be taken. The upper limit $R_{\rm cr}$ is the curve crossing point where $\Gamma(R)$ becomes zero.


\begin{thebibliography}{99}
\bibitem{note}This is in stark contrast to positron-molecule binding which supports very high positron-molecule annihilation rates due to vibrational Feshbach resonances \cite{Gri10}.
\end{thebibliography}
 \end{document}